\documentclass[preprint2]{aastex}
\usepackage{lscape}

\usepackage{url}
\usepackage{amsmath}
\usepackage{color}

\usepackage{footmisc}

\newcommand{\ssst}{\scriptscriptstyle}
\newcommand{\E}[1]{\times 10^{#1}}

\newcommand{\lt}{\left}       \newcommand{\rt}{\right}
\newcommand{\RA}[3]{{#1}^{{\rm h}}{#2}^{{\rm m}}{#3}\fs}
\newcommand{\decl}[3]{{#1}^{\circ}{#2}^{\prime}{#3}\farcs}

\newcommand{\s}{\,{\rm s}}      \newcommand{\ps}{\,{\rm s}^{-1}}
\newcommand{\yr}{\,{\rm yr}}    \newcommand{\Msun}{M_{\odot}}
\newcommand{\cm}{\,{\rm cm}}    \newcommand{\km}{\,{\rm km}}
\newcommand{\kms}{\,{\km\ps}}       
\newcommand{\parsec}{\,{\rm pc}}\newcommand{\kpc}{\,{\rm kpc}}
\newcommand{\erg}{\,{\rm erg}}        \newcommand{\K}{\,{\rm K}}
    \newcommand{\keV}{\,{\rm keV}}
    
\newcommand{\MHz}{\,{\rm MHz}}  \newcommand{\kHz}{\,{\rm kHz}}
\newcommand{\GHz}{\,{\rm GHz}}  \newcommand{\ks}{\,{\rm ksec}}

\newcommand{\Jyperb}{\,{\rm Jy}\,{\rm beam}^{-1}}
\newcommand{\as}{^{\prime\prime}}
\newcommand{\am}{^{\prime}}

\newcommand{\nel}{n_{e}}        \newcommand{\NH}{N_{\ssst\rm H}}
\newcommand{\no}{n_{\ssst 0}}   \newcommand{\rPDS}{r_{\ssst\rm PDS}}
 \newcommand{\tPDS}{t_{\ssst\rm PDS}}

         \newcommand{\vs}{v_{s}}
\newcommand{\nH}{n_{\ssst\rm H}}        \newcommand{\mH}{m_{\ssst\rm H}}
\newcommand{\nHH}{n({\rm H}_{2})} \newcommand{\NHH}{N({\rm H}_{2})}
\newcommand{\VLSR}{V_{\ssst\rm LSR}}
\newcommand{\du}{d_{12}} 
\newcommand{\Eu}{E_{51}}

\newcommand{\twCO}{$^{12}{\rm CO}$}   \newcommand{\thCO}{$^{13}{\rm CO}$}
\newcommand{\eiCO}{${\rm C}^{18}{\rm O}$}
\newcommand{\Jotz}{$J$=1--0}

\newcommand{\Htw}{{\rm H}_{2}}

\newcommand{\xray}{{\rm X-ray}}  
 
 \newcommand{\python}{{\textsc{Python}}}
\newcommand{\XMM}{{\sl XMM}}     \newcommand{\Chandra}{{\sl Chandra}}
\newcommand{\Newton}{{\sl Newton}}  \newcommand{\Fermi}{{\sl Fermi}}
\newcommand{\HESS}{{\sl HESS}}

\newcommand{\EGRET}{{\sl EGRET}} 
\newcommand{\HEASOFT}{{\rm HEASOFT}}

\newcommand{\snr}{{\rm Kes~41}}  \newcommand{\MMSNR}{{\rm thermal composite SNR}}
\newcommand{\HI}{{\ion{H}{1}}}
\newcommand{\gray}{{\rm $\gamma$-ray}}
\newcommand{\nei}{{\sl vnei}}     \newcommand{\phabs}{{\sl phabs}}
\begin{document}
\title{The Metal-Enriched Thermal Composite Supernova Remnant Kesteven 41 
	(G337.8--0.1)
	in a Molecular Environment}

\author{Gao-Yuan Zhang \altaffilmark{1},
Yang Chen \altaffilmark{1,2,6}, 
Yang Su \altaffilmark{3,4}, 
Xin Zhou \altaffilmark{3,4,2}, 
Thomas G.\ Pannuti\altaffilmark{5},
Ping Zhou \altaffilmark{1}}
\altaffiltext{1}{Department of Astronomy, Nanjing University, 163 Xianlin Avenue, Nanjing~210023,
       China}
\altaffiltext{2}{Key Laboratory of Modern Astronomy and Astrophysics,
	    Nanjing University, Ministry of Education, China}
\altaffiltext{3}{Purple Mountain Observatory, CAS, 2 West Beijing Road, Nanjing 210008, China}
\altaffiltext{4}{Key Laboratory of Radio Astronomy, Chinese Academy of Sciences, Nanjing 210008, China}
\altaffiltext{5}{Space Science Center, Department of Earth and Space Sciences, Morehead State University, 235 Martindale Drive, Morehead, KY 40351,
USA.}
\altaffiltext{6}{Author to whom any correspondence should be addressed.}

\begin{abstract}
The physical nature of thermal composite supernova remnants (SNRs) 
remains controversial.  We have revisited the archival \XMM-\Newton\ and
\Chandra\ data of the thermal composite SNR Kesteven~41 (Kes~41 or
G337.8$-$0.1) and performed a millimeter observation 
towards this source in \twCO, \thCO, and \eiCO\ lines.  
The X-ray emission, mainly 
concentrated toward the southwestern part of the SNR, is characterized by
distinct S and Ar He-like lines in the spectra.  The X-ray spectra can be
fitted with an absorbed non-equilibrium ionization collisional plasma model at
a temperature 1.3--2.6~keV and an ionization timescale
$0.1$--$1.2\E{12}\cm^{-3}\s$.  The metal species S and Ar are over-abundant,
with 1.2--2.7 and 1.3--3.8 solar abundances, respectively,
which strongly indicate the presence of a substantial ejecta
component in the X-ray emitting plasma of this SNR.  Kes~41 is found to be
associated with a giant molecular cloud (MC) at a systemic LSR velocity
$-50\km\ps$ and confined in a cavity, delineated by a northern
molecular shell, 
a western concave MC that features a discernible shell, and 
an \HI\ cloud seen toward the southeast of the SNR.
The birth of the SNR in a pre-existing molecular cavity implies a mass
$\ga18\Msun$ for the progenitor if it was not in a binary system.  Thermal
conduction and cloudlet evaporation 
seem to be feasible mechanisms to
interpret the X-ray thermal composite morphology, while the scenario of
gas-reheating by the shock reflected from the cavity wall is quantitatively
consistent with the observations.  An updated list of thermal composite SNRs is
also presented
in this paper.

\end{abstract}


\section{Introduction}
\label{sec:in}
Massive stars rapidly 
evolve to the end of their nuclear-burning lifetimes and explode as core-collapse 
(CC) supernovae before they can move far away from where they formed.
Dozens of CC supernova remnants (SNRs) have been found to be interacting with
molecular clouds (MCs) \citep[see][and references therein]{Jiang2010}.
Such
SNRs are a crucial probe for hadronic interactions between the shock accelerated
protons and the associated MCs,
and many detections of them have been made based on $\gamma$-ray observations
in the current \Fermi\ and \HESS\ era.
Among them are a number of so-called thermal
composite \citep{Jones1998,Wilner1998} or mixed-morphology remnants
\citep{Rho1998}; this class of SNRs is characterized by the coexistence of
radio shells and centrally brightened thermal \xray\ 
emission. The large majority of thermal composite SNRs have been found to be interacting with adjacent MCs \citep{Green1997,Yusef2003}.
Unlike the regular composite type SNRs in which the central \xray\ emission is
clearly known to be powered by pulsars, the nature of the \xray\ kernel in the
thermal composites remains a controversial issue.
\par
Several attempts have been made to interpret the \xray\ morphology of the
thermal composites and 
some mechanisms to explain the center-filled X-ray morphology have been proposed, 
including hot interior
with radiatively cooled rim, thermal conduction in the interior hot gas, gas
evaporation from the shock-engulfed cloudlets, projection effects of the
one-sided shock-cloud interaction, metal enrichment in the interior, shock
reflected inward from 
wind-blown cavity wall, 
amonst other proposed mechanisms \citep[see a summary in][and
references therein]{Chen2008}.  More studies of thermal composite SNRs are
needed to test these mechanisms.
For this purpose,
we here investigate the properties of the interior hot gas and the environs of
SNR Kesteven~41 (\snr\ for the remainder of this paper).
%
%
\par
\snr\ (G337.8$-$0.1), a southern-sky SNR first discovered with the Molonglo
Observatory Synthesis telescope (MOST) at 408$\MHz$ \citep*{Shaver1970}, 
is shown to be centrally
brightened in X-rays within a distorted radio shell by an \XMM-\Newton\
observation and is therefore classified as a thermal composite SNR
\citep{Combi2008}.  In a spectral analysis, the X-ray emission was suggested
to arise from a hot ($\sim 1.4\keV$) gas of normal metal abundance; however,
only the MOS data of the \XMM-\Newton\ have been analyzed so far \citep{Combi2008}.
Presently, the available X-ray observations allow
a more thorough analysis of the physical properties
of \snr.
\par
Moreover, \snr\ has been found to be interacting with an adjacent MC as
indicated by the 1720$\MHz$ hydroxyl radical (OH) maser emission detected on
the radio shell \citep{Koralesky1998,Caswell2004}. The OH satellite line
masers at 1720$\MHz$ are widely accepted as a signpost of SNR-MC interaction
\citep{Lockett1999,Frail1998,Wardle2002}. The SNR-MC interaction may probably
contribute to the high-energy \gray\ excess near \snr\ detected by \EGRET\
\citep{Casandjian2008} and \Fermi\ \citep{Ergin2012} satellites.  The distance
to the remnant is estimated to be 12.3$\kpc$ from the OH maser detection at
the local standard of rest (LSR) velocity $-45\kms$ \citep{Koralesky1998} and
the \HI\ absorption that places it beyond the tangent point at 7.9$\kpc$
\citep{Caswell1975}.
As a follow-up study of the association of \snr\ with an adjacent MC,
we have performed a
millimeter observation toward \snr\ of CO transition lines.
%
\par
In this paper we
carry out an X-ray analysis of SNR~Kes~41 with archival \XMM-\Newton\ (MOS
and PN) and \Chandra\ observation data and investigate the ambient
interstellar environment with our new CO line observation
as well as an archival \HI\ survey. 
\section{Observations and Data Reduction}
\label{sec:observations}

\subsection{\XMM -\Newton\ and \Chandra\ \xray\ Observations}

The \XMM-\Newton\ \xray\ satellite observed \snr\ on September 26th 2005 using
both the EPIC-MOS and the EPIC-PN \citep{Combi2008},
centered at (RA (J2000.0) $=\RA{16}{39}{06}5$, DEC (J2000.0)
$=\decl{-46}{57}{58}0$). The raw EPIC data are cleaned with the latest version
of the Standard Analysis System
(SAS-V13.0.3)\footnote{See http://xmm.esac.esa.int/sas/},
of which the Extended Source
Analysis Software (ESAS) package is mainly used because 
\snr\ appears as an extended source in these observations.
After removing time intervals with heavy proton flarings, the effective 
net exposure time for both the EPIC-MOS and the EPIC-PN is 35$\ks$.
We exclude flux from all the detected point sources 
before performing spatial and spectral analyses.
%
\par
\Chandra\ observed \snr\ with the Advanced CCD Imaging Spectrometer
(ACIS) on June 2011 incidentally in an observation towards high-mass \xray\
binaries in the Norma region (PI: John Tomsick).  There are six sections of
the observation that cover \snr\ region (see Table~\ref{tab:chan_obs} for
observational parameters). We reprocess the events files using the
\Chandra\ Interactive Analysis of Observations (CIAO) data processing software
(Version 4.5)\footnote{See http://cxc.harvard.edu/ciao/}. In the reprocessing, bad
grades are filtered out and good time intervals are reserved.
Flux from all the detected point sources in the field of view (FOV) 
are also removed.

\subsection{CO Observation and Data}

The spectroscopic mapping observation of CO molecular lines, i.e.\
\twCO~(\Jotz) at 115.271$\GHz$, \thCO~(\Jotz) at 110.201$\GHz$, and
\eiCO~(\Jotz) at 109.782$\GHz$, was made with the Mopra 22-m telescope in
Australia on 7--11 July 2011. We did on-the-fly (OTF) mapping with the field of
$\sim 11\am \times 10\am$ covering the 
entire angular extent of \snr.
The 3-mm band receiver and the Mopra spectrometer backend system were used in
zoom mode.
The bandwidth in this mode is about 137.5$\MHz$ (corresponding to about
360$\km\ps$ for \twCO~(\Jotz)) with a spectral resolution of about 33.6$\kHz$
(corresponding to an LSR velocity resolution of $\sim 0.09\km\ps$).
During the observation the system temperature was around 280 K and the
pointing accuracy was mostly better than $\sim 3.5\as$. The half-power beam
size is $\sim 30\as$ at 115$\GHz$, with the main beam efficiency $\sim 0.42$
at the same wave band \citep{Ladd2005}. In this paper, all the images are
presented in antenna temperature,
but  radiative temperature after calibration is used to derive the 
physical parameters of the molecular gas.
\par
The software
\textsc{Livedata}/\textsc{Gridzilla}\footnote{See http://www.atnf.csiro.au/computing/software/livedata/} 
is used to calibrate the
OTF-mapping data and combine them into three-dimensional data cubes,
with two spatial dimensions (longitude/latitude) and one spectral dimension
(LSR velocity). The subsequent data reduction, such as 
subtracting fitted polynomial baselines,
are performed using our own 
script written in \python.
Three data cubes of diverse CO lines are yielded with grid spacing of $30\as$
and velocity resolution of $\sim 0.1\kms$. Finally, the mean rms noise levels of the
main beam brightness temperature are about 0.3, 0.15, and 0.15 K for the
\twCO~(\Jotz), \thCO~(\Jotz), and \eiCO~(\Jotz) lines respectively.

\subsection{Other data}
The archival radio data from the MOST Galactic plane survey at $843\MHz$\
\citep[with an angular resolution of $\sim43\as$;][]{Whiteoak1996}, 
and the SGPS \HI\ data \citep[with an angular resolution of 
$\sim2\am$ and a 
spectral resolution of $\sim0.8\km\ps$;][]{McClure2005} are also
used 
in our multi-wavelength study of \snr.

\section{Analyses and Results}
\label{sec:results}

\subsection{X-ray properties}

\subsubsection{Spatial analysis\label{sec:x_spa}}

In Figure~\ref{fig:xmm} we present a broadband (2.0--7.2~\keV) \xray\ image of \snr\
made from the EPIC-MOS and EPIC-PN data, as
coded
in {\em blue}.
%
In this image,
we have removed the bad pixels, masked the detected point sources 
for both MOS and PN, merged the MOS and PN imaging data with the 
quiescent particle background subtracted and exposure map
correction applied, and adaptively smoothed the image (using the tool 
{\em fadapt} of \HEASOFT\footnote{See http://heasarc.gsfc.nasa.gov/lheasoft/}, version~6.14).
It can be clearly seen that the
\xray\ emission is internally filled without limb brightening as contrasted to
the shell-like feature in radio (shown in white contours).
The centroid of the \xray\ emission sits in
the southwest (SW) region of the northeast-southwest elongated extent of the SNR.
\par
In Figure 2 we present \xray\ images that are produced in the 2.0--2.8~\keV, 
2.8--3.5~\keV\ and 3.5--7.2~\keV\ energy ranges, respectively,
following the similar procedure for the broad band image. 
These images illustrate the photon energy dependence for broadband intensities.
The narrow bands correspond to the sulfur line, argon line (see \S~\ref{sec:xray_spec}),
and higher
energy emission, respectively.
In both the soft (2.0--2.8~keV) and hard
(3.5--7.2~keV) images, there are two \xray\ brightness peaks located in the
geometric center and the southwestern region. 
The hard X-ray image looks
relatively bright in the SW compared
with that in the NE.

All six short sections of the \Chandra\ observation described in
Table~\ref{tab:chan_obs} are also used to generate a merged broadband image of
\snr, but the location of SNR is near to the edge of the CCD chips in each section of
observation, resulting in a decrease of the angular resolution.
The \Chandra\ \xray\ image of the SNR shows no significant
difference from the \XMM-\Newton\ image and is not presented here. 

%
\subsubsection{Spectral analysis}
\label{sec:xray_spec}

Figure~\ref{fig:xray_spec} shows the simultaneous fit of the \XMM-\Newton\ and
\Chandra\ spectra of \snr. The spectral extraction regions for the diffuse emission
source and the local background are defined in Figure~\ref{fig:regions} with
all the detected point sources removed. The spectral extraction is 
such as
to include the 
brightest portion of the diffuse
X-ray emission, which is located in the southwestern part
of the SNR and is
within
the FOVs of all the \XMM-\Newton\ MOS and PN and
\Chandra-ACIS detectors.
Because of the insufficiency of event counts in each spectrum of the
\XMM-\Newton\ EPIC-MOS and \Chandra\ observation, we merge the MOS1 and MOS2
spectra using the \HEASOFT\ tool {\sl addascaspec}, and
merge three \Chandra\ spectra (see Table~\ref{tab:chan_obs}) using {\sl
combine\_spectra} in CIAO 4.5 to increase the statistical quality. 
The other three sections of the \Chandra\ observation hardly cover
the main area of the X-ray emission and are thus not considered.
The spectra are all adaptively regrouped to achieve a background-subtracted
signal-to-noise ratio of 3 per bin. 
In these spectra, there are apparent emission line
features
of metal elements S ($\sim 2.43\keV$) and Ar ($\sim 3.16\keV$), as well as a
hint of Ca ($\sim 3.9\keV$), which confirms the thermal origin of the \xray\
emission.  There is no emission below $\sim2.0\keV$, 
which indicates a large value
of line of sight (LOS) hydrogen column density $\NH$.
\par
The \XMM-\Newton\ and \Chandra\ spectra can be jointly fitted with
an absorbed (using XSPEC model \phabs) single temperature non-equilibrium
ionization (NEI) model (vnei, version~2.0) and the results are summarized in
Table~\ref{tab:joint_fit}.
In the spectral fit, we allow the abundances of S and Ar, which have distinct
line features, to vary and set 
the abundances of the other metal elements equal to solar
\citep{Anders1989}.
The thermal X-ray emitting gas is at a temperature $\sim
1.3$--$2.6\keV$ with an ionization parameter $\nel
t\sim0.1$--$1.2\E{12}\cm^{-3}{\,\rm s}$.
The best fit abundances of the two elements are moderately elevated 
($\sim1.2$--2.7 and $\sim1.3$--3.8 times solar, respectively),
indicating that the hot gas is metal-enriched.

To see where the metal-rich materials' emission arises, we present the
equivalent width (EW) images of S and Ar lines in
Figure~\ref{fig:ewmap}.
These images are constructed using a method similar to that used
in \citet*{Jiang2010SC}, so as to maximize the statistical quality.
Assuming that the 
line-to-continuum 
ratio is intrinsically invariable
for various detectors,
we sum the X-ray counts from \XMM-\Newton\ EPIC-MOS and PN and \Chandra\
ACIS observations band-by-band according to componendo.
We then rebin the data using an adaptive mesh, with each bin in each
narrow band image 
containing 
about
10 counts.
The intensity of the line emission with the interpolated continuum component
subtracted is used as the numerator of the ratio,
and
the background (estimated from the region 
surrounding the SNR) is subtracted from the continuum intensity
which is used as the denominator.
%
Figure~\ref{fig:ewmap} shows that the EW values of both S and Ar lines are
only significant in the interior 
of the southwestern part of the SNR.
The caveat here is that the derived continuum images might have lower
statistical qualities than the corresponding line emission images and not all
the bins can reach 3$\sigma$ significance especially outside the \xray\ emission
region.

%
%

\subsection{The Interstellar Environment}

\subsubsection{Spatial distribution of the ambient clouds}
\label{sec:mc_spa}
\begin{figure}[htpb]
\plotone{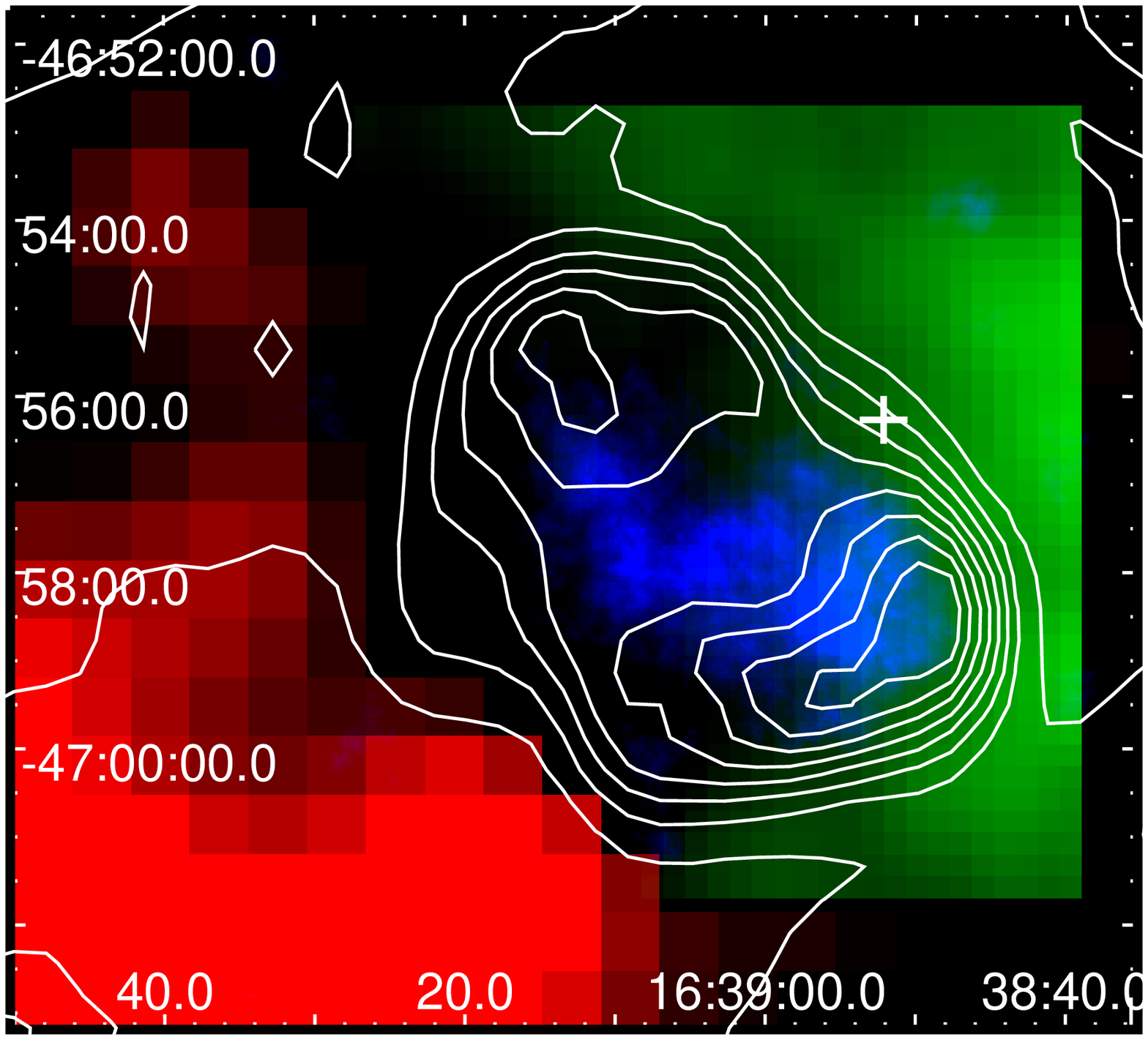}
  \caption{Tricolor image of \snr\ in multiwavelength. Blue: \XMM-\Newton\
  \xray\ image (2.0--7.2$\keV$) obtained by combining the EPIC-PN and MOS data.
  Green: \twCO~(\Jotz) integrated map ($\VLSR=-70$--~$-40\kms$). Red:
  \HI\ line emission from SGPS integrated map ($\VLSR=-55$--~$-50\kms$).
  The image is overlaid with MOST 843$\MHz$ radio contours (at seven linear scale
  levels between 0.00 and 0.79$\Jyperb$; from \cite{Whiteoak1996}). The
  white cross indicates the location of $1720\MHz$ OH maser
  \citep{Koralesky1998} in this SNR.
  }
  \label{fig:xmm}
\end{figure}

The average CO spectra over the FOV are shown in
Figure~\ref{fig:co_spec}. 
There are a number of spectral components along the LOS at the LSR
velocities from
$\VLSR=-130\kms$\ to $\VLSR=0\kms$. 
The LSR velocity of the $1720\MHz$\ maser spot, $\sim -45\kms$,
is 
in the right wing of 
the line component peaked at $\sim-50\kms$ 
which is commonly revealed in the \twCO~(\Jotz), \thCO~(\Jotz), and
\eiCO~(\Jotz)\ spectra (see Figure~\ref{fig:co_spec}).
By an examination of the channel maps of the $\sim-50\km\ps$ line component,
we find a morphological correspondence
between the western radio boundary of the SNR and a 
concave surface of
MC seen in the \twCO\ emission at velocity interval $\sim-55$--$-52\kms$ (see
Figure~\ref{fig:ch_12}).
The concaved MC is also remarkable in the integrated \twCO\ intensity
map in interval $-70$--$-40\km\ps$ within the velocity range of the
line profile (see Figure~\ref{fig:xmm}).
In \thCO\ emission, a bow-like structure at $\sim-50$--$-48\km\ps$ seems to match
the western radio shell;
furthermore, a molecular shell at
$\sim-61$--$-58\km\ps$ perfectly follows the western radio shell of
the SNR (see Figure~\ref{fig:ch_13}).
In addition, a section of molecular shell in \twCO\ emission at
$-47.5\km\ps$ is seen to closely follow the northern radio shell (see
Figure~\ref{fig:ch_12}). 
Such correspondence is actually only seen in this spectral component by
examining the data cubes across the entire LSR velocity span. 
Around the OH maser position, we can also see some dense molecular gas
in the intensity maps (Figure~\ref{fig:ch_18}) of \eiCO\ emission,
which is optically thin and traces molecular cores.
The combination of the OH maser and the spatial features of CO emission
presented here clearly demonstrates that SNR \snr\ is in physical
contact with the MC at a systemic velocity $\VLSR\sim-50\km\ps$ in the west and the 
north,
although the broadened CO line profiles of the $\sim-50\km\ps$ molecular component
due to shock disturbance 
can not be discerned because of line crowding along the LOS.

\par
It is noteworthy that a dense \HI\ cloud at $\VLSR\sim-52\km\ps$ is situated
to the southeast of the SNR  (see Figure~\ref{fig:xmm}, where the \HI\
intensity image is presented in velocity interval $-55$--$-50\kms$),
and
the southeastern radio shell of the SNR seems to follow the surface of the \HI\ 
cloud.
SNR~\snr\ appears to be 
confined in a cavity enclosed by the molecular cloud in the west and northwest
and the \HI\ cloud in the southeast.
The archival data show that this \HI\ cloud has an angular size of 
$\sim18\am$.
%
\subsubsection{Parameters of the ambient clouds}\label{sec:mc_para}

We estimate the gas parameters of the $-50\km\ps$ molecular component
in velocity range $-70$--$-40\kms$ 
with which the SNR is interacting,
and the results are listed in Table~\ref{tab:mc}. 
The column density and 
mass of the molecular gas derived may be an
upper limit, 
because there may be irrelevant contributions in 
the complicated line profiles in \twCO\ and \thCO\ emission.
For the estimate, we adopt a box region including the main part of the MC in
the FOV (see the $-53.5\km\ps$ panel of Figure~\ref{fig:ch_12}).
Three methods have been utilized to estimate
the average molecular column density $\NHH$, which give a similar
value $N(\Htw)\sim3$--$5\E{22}\cm^{-2}$. In the first
method, we use the mean CO-to-$\Htw$ mass conversion factor (known as the X-factor)
1.8$\E{20}\cm^{-2}\K^{-1}\km^{-1}\s$ \citep{Dame2001} to calculate the molecular
column density based on the \twCO~(\Jotz) emission.
In the second method, the \thCO\ column density is converted to the $\Htw$
column density using
$N(\Htw)/N($\thCO$)\approx6.2\E{5}$ \citep{Nagahama1998}, and in the third
method, the relation $N(\Htw)/N($\eiCO$)\approx7\E{6}$ \citep{Warin1996} is used for the conversion
from \eiCO\ column density to $\Htw$ column density.
In the second and the third method, we have assumed a local thermodynamic equilibrium and \twCO~(\Jotz) line being optically thick,
with an excitation temperature of 12.9$\K$ (as obtained from the line peak of
\twCO~(\Jotz)).

Prior to deriving the column
densities of \thCO\ and \eiCO, we have derived the optical depths
$\tau($\thCO$)\approx0.6$ and $\tau($\eiCO$)\approx0.1$,
and therefore we treat the \thCO\ and \eiCO\ emission as optically thin.
The $\NHH$ value obtained from the third method,
$\sim5.1\E{22}\cm^{-2}$,
is a little greater than those obtained from the other two
methods, and it is probably because \eiCO\ traces denser clouds.
\par

The archival data show that the \HI\ cloud southeast to \snr\ also has a
spectral peak at $-50\km\ps$ and similar velocity span to the spectral profile
of the $-50\km\ps$ \twCO\ component.
The column density of \HI\ gas ($T_B\sim110\K$) is estimated 
to be $\sim5.1\E{21}\cm^{-2}$, using the relation $N$(\HI)=1.823$\E{18}T_B \Delta v$ 
on the optically thin assumption
\citep*{Dickey1990}.

\section{Discussion}\label{sec:dis}
\subsection{SNR environment}

Our analysis has revealed that SNR~\snr\ is confined in a cavity 
enclosed by MCs in the west and the north as well as an \HI\ cloud in the 
southeast (\S~\ref{sec:mc_spa}). Some shell-like features of molecular gas also appear to 
follow the periphery of the SNR.
 We can give some crude estimates on the gas density of the complicated surrounding environment.

Assuming the LOS size of the $\sim-50\kms$ MC is similar to its apparent size $\ga 9\am$
(a lower limit given by the FOV) and using the molecular column density
$\NHH\sim3$--$5\E{22}\cm^{-2}$, we estimate the molecular
density as $n(\Htw)\la310$--$510d_{12}^{-1}\cm^{-3}$, where $d_{12}=d/(12 \kpc)$ is
the distance to the MC associated with the SNR in units of the reference value
estimated from the maser observation \citep{Koralesky1998}.
\par
Another density estimate can be made from the excavation of the molecular gas
in the SNR extent, following the method used in \citet{Jiang2010}.
We plot the molecular column density distributions of the \twCO\ line complex
in interval $-70$--$-40\kms$ along the two diagonals of the FOV 
using X-factor method (see Figure~\ref{fig:col}).
As can be clearly seen,
$N(\Htw)$ generally increases both
from the southeast to
the northwest and from the northeast to the southwest.
Notably, there are peaks/bumps in column
density just outside the
SNR boundary, which is consistent with the
molecular shells described in \S~\ref{sec:mc_spa}.
Outside the bumps of the both $\NHH$ curves, there are platforms
at $\sim4\E{22}\cm^{-2}$;
just inside the opposite boundaries, there are lower platforms (wide in the {\it
left} panel and narrow in the {\it right} panel) both at $\sim2.9\E{22}\cm^{-2}$.
The common difference of column density $\sim 1.1\E{22}\cm^{-2}$ can be
explained with the removal of molecular gas from the SNR region. 
%
Thus the molecular gas originally in the excavated region
has a number density 
$n(\Htw)\sim140d_{12}^{-1}\cm^{-3}$ assuming the mean LOS length of
the excavated region is similar to the minor axis of the elongated SNR.

If the apparent size of the southeastern \HI\ cloud --- that is, $\sim18\am$ (see
\S~\ref{sec:mc_spa}) --- is adopted as the LOS size of the cloud, then the 
number density of the neutral gas can be estimated from the neutral column density as
$n$(\HI)$\sim 36\du^{-1}\cm^{-3}$.
In the \xray\ spectral fitting, a high hydrogen column density of
$\sim5.5$--$8.0\E{22}\cm^{-2}$ is needed to explain the absence of soft ($\la2.0\keV$) \xray\
emission. 
Apparently, the \HI\ gas towards the \snr\ direction, which has a
column density $\sim2\E{22}\cm^{-2}$
\citep{Kalberla2005,Dickey1990}
\footnote{See http://heasarc.gsfc.nasa.gov/cgi-bin/Tools/w3nh/w3nh.pl}, 
is not sufficient to account for the X-ray absorption.
However, the $\VLSR\sim-50\km\ps$ MC, which is associated with the SNR
and seems to dominate 
the CO emission in the direction (Figure~\ref{fig:co_spec}),
has a hydrogen column density $\la7$--$10\E{22}\cm^{-2}$
and can significantly contribute to the absorbing column.
Moreover,
the harder X-ray emission in the SW than in the NE (\S~\ref{sec:x_spa})
may be consistent with the heavier absorption by the denser 
molecular gas located in the west.

\subsection{SNR physics}


\subsubsection{Comparisons with main mechanisms and models}
\label{sec:confronted}
\emph{(i) Some physical parameters.}
The density of the X-ray emitting gas interior to the SNR can be 
estimated using the volume emission measure obtained from the spectral
analysis.
On the assumption $\nel\approx1.2\nH$ for the densities of electrons and 
hydrogen atoms, we get
$\nH\sim0.2$--$0.4f^{-1/2}\du^{-1/2}\cm^{-3}$, where $f$ is the filling factor
of the hot gas.
We estimate the ionization age $t_i\sim4\E{3}$--$1.1\E{5}f^{1/2}\du^{1/2}\yr$
by using the best-fit ionization parameter
$\nel t$ (also see Table~\ref{tab:joint_fit}).

In the X-ray spectral analysis, the spectral extraction region covers the
main part of the X-ray emission.
The mass of the hot gas in the region is
$\sim2$--$4f^{3/2}\du^{7/2}\Msun$, which can be regarded as a lower limit.
If the hot gas pervades the whole interior of the remnant,
we obtain an upper limit of the mass of the
hot gas, $\sim18$--$37 f^{1/2}\du^{7/2}\Msun$,
assuming a prolate ellipsoid volume
($12.6\times8.4\times8.4\du^{3}\parsec^3$) for the remnant.

\emph{(ii) Sedov evolution.}
\snr\ has been shown to be surrounded by molecular gas of density
$\nHH\sim140$--$350\cm^{-3}$ and \HI\ gas of density
$n(\mbox{HI})\sim40\cm^{-3}$.
If the SNR follows the \citet{Sedov1959} evolution in the dense ambient gas,
and the hot gas temperature obtained from the X-ray spectral fit, $kT_x$,
is taken to be the emission-measure weighted mean temperature, then
the postshock temperature would be $kT_s=kT_x/1.27$ and the expansion
velocity would be $v_s=(16kT_s/3\bar{\mu}\mH)^{1/2}\sim1.1^{+0.2}_{-0.1}\E{3}\km\ps$,
where the mean atomic weight $\bar{\mu}=0.61$ is used.
The dynamic age would be $t=2r/5v_s\sim5.8^{+0.8}_{-1.3}\E{3}\du\yr$ (where $r\sim13\du\parsec$ is the mean radius),
and the supernova explosion energy would be
$E=(25/4\xi)1.4\no\mH r^3\vs^2\sim2\du^3\E{53}\erg$ or $2\du^3\E{54}\erg$,
for ambient density $\no=40\cm^{-3}$ (for \HI\ gas) and $280\cm^{-3}$
(for molecular gas), respectively, where $\xi=2.026$.
Such estimates are highly unphysical, because the explosion energy
is much higher than the canonical value $10^{51}\erg$. Moreover,
the molecules in the surrounding molecular shell (\S~\ref{sec:mc_spa})
could not survive
if the shell were swept up by the SNR shock, given the high shock velocity.
Therefore, the assumption of Sedov evolution of \snr\ in a uniform medium is problematic.

\emph{(iii) Evolution in radiative phase.}
The radiatively cooled rim is one of the scenarios invoked to explain the
X-ray thermal composite morphology.
If \snr\ is not considered to be in a pre-existing cavity,
then
the radiative pressure-driven snowplow (PDS) stage begins at a radius
\citep{Cioffi1988}
\(
\rPDS=2.9\Eu^{2/7}(\no/40\cm^{-3})^{-3/7}\zeta_{m}^{-1/7}\parsec
\)
and an age
\(
\tPDS = 3.6\E{3}\Eu^{3/14}(\no/40\cm^{-3})^{-4/7}\zeta_{m}^{-5/14}\yr
\),
where $\Eu\equiv E/(10^{51}\erg)$ and 
$\zeta_{m}$ metallicity factor of order unity. 
If $\no=280\cm^{-3}$ for the detected molecular gas, then
$\rPDS\sim1.3\parsec$ and $\tPDS\sim1.8\E{3}\yr$.
Compared to the mean radius $\sim13\parsec$ and the ionization
age $\ga4$~kyr,
these estimates indicate that the SNR may have entered the radiative phase
at an early time.
In this scenario, the present velocity of the blast wave is
\(
\vs=21[\Eu(\no/40\cm^{-3})^{-36/31}\zeta_{m}^{-5/31}(r/13\parsec)^{-98/31}]^{31/42}\km\ps
\)
\citep{Cioffi1988,Chen2001},
which would be $\sim4\km\ps$ for an ambient molecular gas with $\no=280\cm^{-3}$,
where $\Eu\sim1$ and $\zeta_m\sim1$ are assumed
for simplicity.
With such a shock velocity, the molecular shell can plausibly be 
explained to be the material swept up by the SNR blast wave.
The dynamic age of the remnant is thus given by 
\(
	t=3.3\E{3}(r/14\parsec)
		  (\vs/413\km\ps)^{-1}
			[3 + \no^{-10/31} \zeta_{m}^{-10/31}
			  (r/14\parsec)^{-10/31}
				(\vs/413\km\ps)^{40/31} ]\yr
\)
\citep{Cioffi1988,Chen2001},
which yields $t\sim1.8\E{5}\yr$ for a preshock atom gas with
$\no\sim40\cm^{-3}$ or 
$1.0\E{6}\yr$ for a molecular gas with $\no\sim280\cm^{-3}$,
respectively, where $r=13\parsec$ is adopted.
The age estimates here, however, appear to be larger than the ionization
age, $t_i\sim4$--$110f^{1/2}\du^{1/2}$~kyr.


\emph{(iv) Thermal conduction.}
Thermal conduction may play a role in shaping the centrally brightened
thermal X-ray morphology of \snr\ by lowering the temperature
and increasing the density of the hot gas in the central portion.
If the suppression of conduction by magnetic fields can be ignored,
the
conduction timescale of the hot gas inside the remnant is given by $t_{cond}\sim n_e k r^2/\kappa$, where
the conductivity is given by
$\kappa=1.8\E{-5}T_x^{5/2}/\ln\Lambda$, with the Coulomb logarithm
$\ln\Lambda=29.7+\ln n_e^{-1/2}(T_x/10^6\K)$ \citep{Spitzer1962}.
The timescale would thus be
$t_{cond}\sim1.5$--$3.1$~yr, shorter than the ionization age of
the X-ray emitting gas.

\emph{(v) Cloudlet evaporation.}
As another popular scenario for thermal composites, cloudlet evaporation
is possibly taking place in the remnant interior,
since \snr\ is evolving in an interstellar environment with
molecular gas which is usually clumpy and dense clumps can be engulfed
by the blast wave.
In this case, hot gas with a mass up to $\sim18$--$37\Msun$ is mostly
composed of evaporated gas. This mass is not high enough to substantially dilute
the supernova ejecta of a few solar masses, and therefore the evaporation
scenario may not be in conflict with the detected metal enrichment.
Limited X-ray counts available in the archived observations prevent
an estimate of the spatial distribution of such physical parameters
as the temperature and density of the hot gas.
Even the X-ray surface brightness shows severe asymmetry primarily due to
the presence of two brightness peaks (\S~\ref{sec:x_spa}), which also makes it difficult
to locate the explosion center.
All these factors hamper us
from presenting a quantitative comparison between the observational
properties and the evaporation effect predicted in the \citet{White1991}
model.



\subsubsection{Metal enrichment}
Metal enrichment in the thermal composite SNRs complicate the
explanation of the physical mechanism for this class, but meanwhile it is
also an important clue to understanding this class of sources.
\snr\ is a thermal composite SNR
where the abundances of a few metal species (S and Ar) are elevated.
This result motivates its classification as
new member of the subclass of
``enhanced-abundance'' 
\citep{Lazendic2006} or ``ejecta-dominated''
\citep{Pannuti2014} thermal composite SNRs.
This subclass has members amounting to a number of 25--30 (see an updated list
of thermal composite SNRs in Table~\ref{tab:mm}), 
comprising
about 67\%--81\% of the 36--37
known thermal composites.
Sulfur and silicon are the most common enriched species seen among these sources:
they are overabundant in 18--19 and 14--15 such SNRs, respectively.
The inert elments such as Ar and Ne are overabundant in thirteen thermal composites.

In the case of SNR~W44, \citet{Shelton2004} suggest
that central metal enrichment is a factor
as important
as thermal conduction for the explanation
of the centrally-brightened X-ray emission from thermal composite SNRs.
They interpret the enhanced metal abundance of W44 
as originating from
destroyed
dust grains within the SNR as well as ejecta enrichment.
In \snr, the dust destruction scenario may not be favorable because 
argon, obviously enriched, is an inert element and cannot easily
depleted into dust.
In the most recent X-ray studies of G352.7$-$0.1 and W51C,
metal enrichment is explained to come from supernova ejecta
\citep{Pannuti2014,Sasaki2014}.


As pointed out by \citet{Vink2012}, metal-rich plasma produces more X-ray
emission than plasma that is not metal enhanced.
In \snr,
the S and Ar emissions contribute 7\% and 2\%, respectively, to the
broad energy ranges $0.5$--$10\keV$, and the extra abundances of these two
metals contribute 3\% and 1\%, respectively.

\subsubsection{Effect of pre-existing cavity}
\label{sec:cavity}
\emph{(i) Shock reflection.}
Thermal composite SNRs are usually associated with dense material 
(especially MCs), and therefore their
progenitor stars are likely to have created a cavity in the dense
environment with energetic stellar winds and ionizing radiation.
Actually, about ten thermal composites have been found or suggested
to evolve in cavities (see Table~\ref{tab:mm}).
When the supernova blast wave hits the cavity wall, a reflected
shock can be expected to be sent back inwards and reheat the interior gas
including the ejecta to a higher temperature.
This is the case suggested for the thermal X-ray composite
morphology to SNR Kes~27 \citep{Chen2008}. A similar explanation is also proposed
for the thermal composite SNR 3C~397, which is embedded in the edge of a giant MC and confined by a
sharp molecular wall \citep{Jiang2010}. 
The thermal composite SNR W44
appears to be confined in a molecular cavity that is
delineated by sharp molecular ``edges" in the east, southeast, and 
southwest, as well as molecular filaments and clumps in other 
periphery positions \citep{Seta2004}.
The metal-enriched X-ray emitting gas in the interior of the SNR may
have been reheated by the shock reflected from the
cavity wall (which is at least represented by the sharp ``edges"). 

\snr\ has been demonstrated here to also be confined in a cavity
in a molecular environment.
As shown above (\S~\ref{sec:confronted}), both the Sedov and radiative-phase
evolution in a uniform dense molecular/\HI\ gas
will lead to unreasonable physical parameters, and hence the cavity
is not blown by the SNR itself.
A collision of the SNR shock with the pre-existing cavity wall thus
cannot be ignored.
The forward shock of the SNR is drastically decelerated upon
the collision. Assuming that the density ratio between
the cavity and the wall is much smaller than 1 and adopting
the cavity radius $r_c\sim13\parsec$, the shock velocity upon the collision
can be essentially 
estimated as \citep{Chen2003}
\begin{equation}
v_s(r_c)=\lt[\frac{E}{2\pi(1.4\no\mH)r_c^3}\rt]^{1/2},
\end{equation}
which is $\la20\Eu^{1/2}$ and $50\Eu^{1/2}\km\ps$ for $\no\ga280\cm^{-3}$ (for
molecular gas) and $\no\sim40\cm^{-3}$ (for \HI\ gas), respectively.
Such low velocities indicate that \snr\ has entered radiative
stage right after the blast wave struck the cavity wall.
This effect readily explains the lack of X-ray emission in the
SNR rim.

The interior gas remains hot and is elevated to a higher temperature by
the heating by a reflected shock. This scenario can be tested in combination
with the observation result and simplified theory presented by \citet{Sgro1975}.
For a shock reflection at a dense wall, the temperature ratio between 
the post--reflected-shock gas and the post--transmitted-shock gas is given by
\begin{equation}
\frac{T_r}{T_t}=\frac{A}{A_r}, \label{eq:Tratio}
\end{equation}
where $A_r$ is the density contrast between the post--reflected-shock
gas and
the post--incident-shock gas and $A$ is the wall-to-cavity
density contrast, which in turn is a function of $A_r$, specifically,
\begin{equation}
\label{equ:ar}
A=\frac{3A_r(4A_r-1)}{\{[3A_r(4-A_r)]^{1/2}-\sqrt{5}(A_r-1)\}^2}.
\end{equation}
We take $\nHH\sim140\cm^{-3}$ as a typical density of the molecular gas
and adopt the velocity of the transmitted shock in the molecular cavity wall
to be $\sim20\km\ps$, which corresponds to $kT_t\sim2.2$~eV.
If the observed temperature of the hot gas, $T_x$, is adopted for $T_r$,
we can calculate the dependence of the wall-to-cavity density $A$ on
$T_r$ in the uncertainty range 1.3--2.6~keV.
Figure~\ref{fig:ref} shows that $A$ is in the range of 1500--3000.
Thus a cavity density $\sim0.1$--$0.2\cm^{-3}$ is implied.
If we adopt the observed upper limit $500\cm^{-3}$ for $\nHH$, 
almost the same cavity density is inferred.
Such a cavity density seems to be of order similar to the observed
hot gas density.
Here, shock reflection from the dense molecular wall is more efficient
than from the \HI\ cloud; actually,
the centroids of both the diffuse X-ray emission (\S~\ref{sec:x_spa}) and the EW maps
of S and Ar (\S~\ref{sec:xray_spec}) in the southwestern half can be consistent with
the reheating primarily by the shock reflected from southwestern dense MC.



\emph{(ii) Progenitor's mass.}
The pre-existing cavity of \snr\ should be excavated by the progenitor
star, and the progenitor's mass can be estimated from the cavity size.
A linear relation has been found between the progenitor's mass, $M_{prog}$,
and the radius, $R_b$, of the bubble blown by the main-sequence wind
in a giant MC \citep{Chen2013}:
\begin{equation}
R_b\approx (1.22\pm0.05)(M_{prog}/\Msun)-(9.16\pm1.77)\parsec.
\end{equation}
If we adopt $R_b\approx r_c$, the mean radius of the cavity, a stellar
mass $18\pm2\Msun$ is obtained. However, the cavity is not purely
confined by molecular gas, but by a giant \HI\ cloud in the southeastern
side, and there seems to be a blow-out morphology in the northeast
(see Figure~\ref{fig:xmm}). Therefore, this mass estimate should be regarded as
a lower limit of the progenitor's mass if the progenitor was not in a binary
system, which implies that the progenitor's
spectral type is no later than B0.

\emph{(iii) Ejecta in cavities.}
It can be seen in Table~\ref{tab:mm}, about half out of the 36-37 known thermal
composite SNRs are interacting with adjacent MCs; moreover, 
eleven or twelve of them 
are known/thought to be evolving in cavities in dense gas,
and most of them exhibit overabundance of metal species.
This fact can be explained in the context of the ejecta in pre-existing
cavities.
An appreciable fraction of supernova ejecta, especially some heavy metal species like Si, are expected to expand
at a relatively low velocity \citep[e.\ g., $\sim 10^3\km\ps$; e.\ g.,][Mao.\ J.\ 
et al.\ in preparation]{Ono2013};
for the ejecta expanding in pre-existing cavities, the slow part
may have not been substantially mixed or diluted by the interstellar
materials by the time when the blast shock is reflected backwards
from the cavity wall.
They are not only heated by the reverse shock in the early free expansion phase, but also can 
very possibly be reheated by the reflected shock.
Therefore, the hot interior are thus observed to
be metal enriched in X-rays.

\subsubsection{Recombining effect?}
A number of thermal composite SNRs have been revealed to contain
recombining plasma (see the rows labeled with ``\emph{R}'' in Table~\ref{tab:mm}).
However, the over-ionization in the case of \snr\ can be neither
proven nor disproven by the spectral fitting due to the lack of obvious radiative
recombination continuum (RRC) features and H-like lines of the same metal species
in the X-ray spectra
with insufficient counts. 
If there is recombining plasma in the remnant,
the temperature obtained here using the {\sl vnei} model may be an overestimate
in view of the ``apparent'' hard continuum raised by the RRC. Deeper X-ray
observations are needed 
to determine if the X-ray-emitting plasma of \snr\ is indeed overionized.

\section{Summary}
We have carried out an X-ray observational analysis by revisiting the 
archival \XMM-\Newton\ and \Chandra\ data of the thermal composite
SNR~\snr. We have also conducted and analyzed an observation
towards the remnant in \twCO, \thCO, and \eiCO\ lines.
The main results are summarized as follows.
\begin{enumerate}
\item Thermal X-ray emission from \snr\ is detected above 2~keV,
which is characterized by distinct S and Ar He-like lines and a
brightening in the SNR interior, with a centroid in the emission toward the 
southwest portion of the SNR.
\item The X-ray emitting gas can be described as an optically thin
plasma (at a temperature 1.3--2.6~keV, with an ionization parameter
$0.1$--$1.2\E{12}\cm^{-3}\s$) with a NEI model.
The metal species S and Ar are found to be over-abundant, with
abundances 1.2--2.7 and 1.3--3.8 solar, respectively.
They may be enriched by the supernova ejecta.
\item The SNR is found to be associated with a giant MC at a systemic
LSR velocity $-50\km\ps$ and is confined in a cavity 
delineated by a northern molecular shell, western concave MC with
also a discernable shell, and a southeastern \HI\ cloud.
\item The cavity is pre-existing prior to the supernova explosion;
it would not be physically realistic to conclude that it was generated
by the SNR expansion.
\item \snr\ seems to have left the adiabatic stage of evolution and 
entered the radiative phase with an \xray\ dim rim just after the
SNR shock encountered the dense interstellar medium cavity.
\item Thermal conduction and cloudlet evaporation seem to be
feasible mechanisms to interpret the X-ray thermal composite
morphology, while the scenario of gas reheating by shock reflected
from the cavity wall is quantitatively consistent with the observations.
\item The birth of the SNR in a cavity in a molecular environment
implies a mass $\ga18\Msun$ for the progenitor if it was a single star.
\end{enumerate}

\begin{acknowledgements}
G.Y.Z.\ is grateful to the COSPAR Capacity Building Workshop, 2013 and Randall
Smith for the advice on the practical skills of X-ray analysis.  Y.C., Y.S.,
and X.Z.\ thank the support of NSFC grants 11233001, 11103082, and 11403104.
This work is also beneficial from the 973 Program grant 2015CB857100,
the grant 20120091110048 by the Educational
Ministry of China and the grant BK20141044 by Jiangsu Provincial Natural Science Foundation.
This research has made use of NASA's Astrophysics Data
System and pywcsgrid2\footnote{http://leejjoon.github.com/pywcsgrid2/}, an open-source plotting packge for Python.

\end{acknowledgements}

\bibliographystyle{apj}
\bibliography{cite}

\begin{deluxetable}{cccccccc}
  \tablecaption{Archival \Chandra\ observations for \snr\ region}
  \tabletypesize{\footnotesize}
  \tablecolumns{8}
  \tablehead{ &&& \colhead{Observation}& \colhead{Exposure}& \colhead{CCD}&
  \colhead{Good Exposure}& \colhead{Spectral}\\
  \colhead{Obs.\ ID}  & \colhead{R.A.} & \colhead{Decl.} & \colhead{Date} &
  \colhead{($\ks$)} & \colhead{chips \tablenotemark{a}} & \colhead{($\ks$)} &
  \colhead{ analysis \tablenotemark{b}} \\
  }
	\startdata
	12513 & $\RA{16}{39}{53}81$ & $\decl{-46}{57}{45}0$ & 06/27/11 & 20.4 & I0 & 20.2 & yes \\
	12514 & $\RA{16}{39}{01}38$ & $\decl{-46}{49}{45}8$ & 06/10/11 & 20.0 & I1 & 19.8 & no  \\
	12516 & $\RA{16}{39}{06}95$ & $\decl{-46}{06}{42}5$ & 06/11/11 & 19.8 & I2 & 19.5 & no  \\
	12517 & $\RA{16}{38}{14}49$ & $\decl{-46}{58}{42}0$ & 06/11/11 & 19.8 & I3 & 19.5 & yes \\ 
	12519 & $\RA{16}{38}{19}83$ & $\decl{-47}{15}{38}8$ & 06/13/11 & 19.6 & S2 & 19.3 & no  \\
	12520 & $\RA{16}{37}{27}34$ & $\decl{-47}{07}{37}0$ & 06/13/11 & 19.6 & S3 & 19.0 & yes \\
    \enddata
  \tablenotetext{a}{Indicates the chips which sampled emission from \snr.
  }
  \tablenotetext{b}{Indicates which observations featured data that were extracted for spectral analysis.
  }
  \label{tab:chan_obs}
\end{deluxetable}

\begin{deluxetable}{lc}
\tablecaption{Spectral fit results with 90\% confidence range for the
X-ray emitting gas in SNR~\snr \label{tab:joint_fit}}
  \tablewidth{0pt}
  \tablehead{
  \colhead{Parameter}  & \colhead{Value} } 
  \startdata
    Net count rate ($10^{-3}\,$counts$\,\s$)  &  $28.05\pm1.60$\tablenotemark{a}\\
	 & $7.86\pm0.49$\tablenotemark{b} \\
	 & $2.46\pm0.38$\tablenotemark{c} \\
	$\NH$($10^{22} \cm^{-2}$)    & 6.75$^{+1.27}_{-1.21}$\\
	$kT_x$(\keV)                    & 1.80$^{+0.82}_{-0.47}$ \\ 
	$\nel t$($10^{11}\s$ $\cm^{-3}$)     & 2.06$^{+9.87}_{-1.56}$\\
	$f \nel \nH V/d_{12}^2 $($10^{57}$cm$^{-3}$)
	& 1.46$^{+1.37}_{-0.72}$\\{}
	[S/H]         & 1.80$^{+0.92}_{-0.65}$              \\{} 
	[Ar/H]        & 2.40$^{+1.42}_{-1.06}$              \\ 
	Reduced $\chi^{2}$ (d.o.f.) & 0.99 (54)             \\
	Flux\tablenotemark{d} ($10^{-12}\,$erg\,$\cm^{-2}\,{\rm s}^{-1}$) & 1.9 \\
	$\nH f^{1/2} d_{12}^{1/2}$\tablenotemark{e} ($\cm^{-3}$) & $0.3\pm0.1$\\
	$t_i f^{-1/2}\du^{-1/2}\,(10^{4}\yr)$ & $1.9^{+8.9}_{-1.5}$
  \enddata
  \tablenotetext{a,b,c}{The count rates are from PN, MOS, and ACIS, respectively,
  with $1\sigma$ error range.}
  \tablenotetext{d}{The unabsorbed fluxes are in the 0.5--10$\keV$ band.}
  \tablenotetext{e}{In the estimate of the densities, we assume a volume of 
  a prolate ellipsoid (considering the NE-SW elongated morphology of the SNR) with size $5.8\times4.2\times4.2\du^3\parsec^{3}$ for 
  the elliptical region of \xray\ spectral extraction 
  (see Figure~\ref{fig:regions}).}
  \tablecomments{The abundances of all elements are set to
  solar abundances except for S and Ar.}
\end{deluxetable}

\begin{deluxetable}{ccccc}
\tablecaption{Derived Parameters of the Ambient Clouds in Velocity
Interval $-70$--$-40\kms$ \label{tab:mc}}
  \tablewidth{0pt}
  \tablehead{
  \colhead{Lines}  & \colhead{$T_{\rm peak}$(K)\tablenotemark{a}} &
  \colhead{$N(\Htw)\,(10^{22}\cm^{-2})$} &
  \colhead{$M(10^4\Msun)\tablenotemark{b}$}&
  \colhead{$\tau$}}
  \startdata
	\twCO    & 9.5 &  3.8 &
	21.1$d_{12}^{2}$
	&- \\
	\thCO    & 4.5 &  3.5    & 19.5$d_{12}^{2}$ &0.64 \\
	\eiCO    & 0.95&  5.1    & 28.3$d_{12}^{2}$ &0.10 \\
  \cline{3-3}
 & & $N($\HI$)\,(10^{21}\cm^{-2})$ & 
 & \\
  \cline{3-3}
	\HI \tablenotemark{c}     & 110 &  5.1& 
	1.9$d_{12}^{2}$ &- \\ 
  \enddata

  \tablenotetext{a}{The average peak temperature in the box region decipted in Figure~\ref{fig:ch_12}, corrected to radiative temperature already.}
  \tablenotetext{b}{$d_{12}=d/(12 \kpc)$ is
the distance to the MC associated with the SNR in units of $12\kpc$.}
  \tablenotetext{c}{We use the same LSR velocity range
  $-70$--$-40\kms$ for the HI gas, and this range likewise includes the main
  part
  of the HI line profile peaked at $-50\km\ps$.
}
  
\end{deluxetable}
\onecolumn

\begin{figure}[htpb]
\plotone{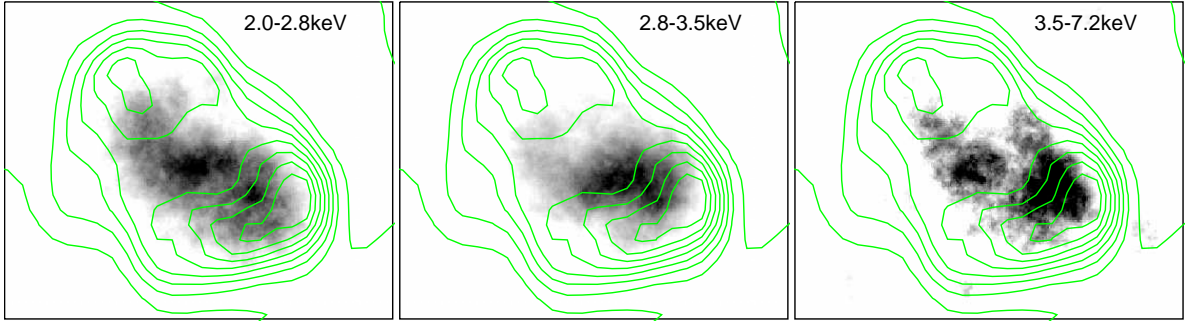}
  \caption{The \XMM-\Newton\ (MOS+PN) X-ray images of Kes~41 in energy bands 2.0-2.8$\keV$,
  2.8-3.5$\keV$, and 3.5-7.2$\keV$, with radio contours (the same as in
  Figure~\ref{fig:xmm}) overlaid.  }
  \label{fig:tri_x}
\end{figure}

\begin{figure}[htpb]
\includegraphics[angle=270,width=\columnwidth]{same_bk.eps}
  \caption{\xray\ spectra of the region defined in Figure \ref{fig:regions}
  from both \XMM-\Newton\ and \Chandra\ observations. Red points show the EPIC-PN data; the
  black points are from the merged EPIC-MOS1/2 data; the green points show the
  co-added spectrum made by merging all 3 \Chandra\ observations that cover the
  majority of the SNR's \xray\ emission
  as described in Section \ref{sec:xray_spec}. All the
  spectra have been adaptively binned to achieve a background-subtracted S/N
  (signal-to-noise ratio) of 3 per bin.
  They are jointly fitted using the same absorbed
  \nei\ model and the lines with different colors are fitted model lines matching
  the same color data points separately.} \label{fig:xray_spec}
\end{figure}

\onecolumn
\begin{figure}[htpb]
\plotone{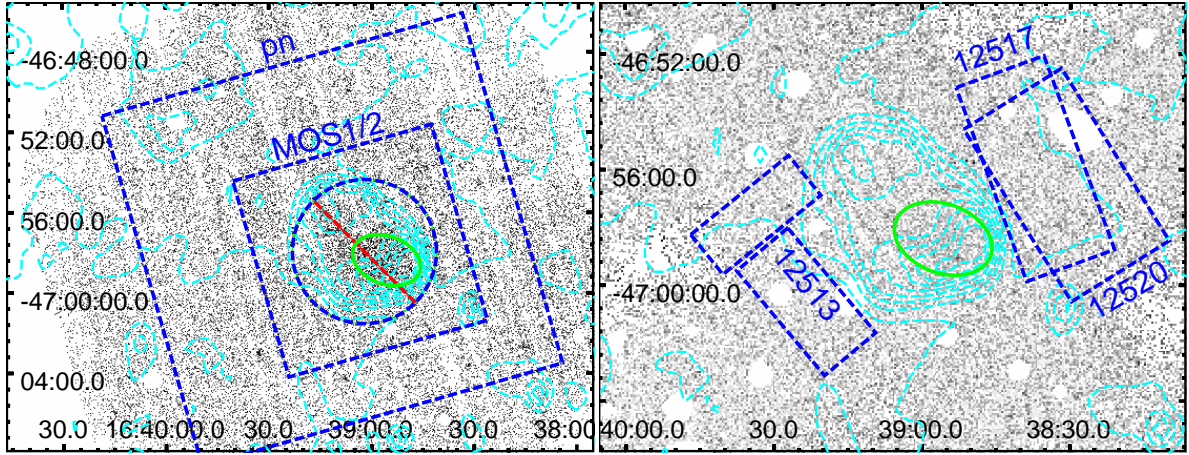}
  \caption{Raw images of the \XMM-\Newton\ observation (left) and merged
  \Chandra\ observations (right), with all detected point sources removed.
  The common ellipse (in green) is used for source spectral extraction. 
  In the left panel, the dashed boxes, excluding the dashed circle with the
  crossing 
  line (in red), are used for PN and MOS1/MOS2 background
  subtraction.
  In the right panel, the boxes with the numbers
  denoting the Obs.\ IDs in Table~\ref{tab:chan_obs} are used as background
  regions for the corresponding data. The cyan contours are from radio observation,
  with levels
  the same as in Figure~\ref{fig:xmm}.} \label{fig:regions}
\end{figure}

\begin{figure}[htpb]
\plotone{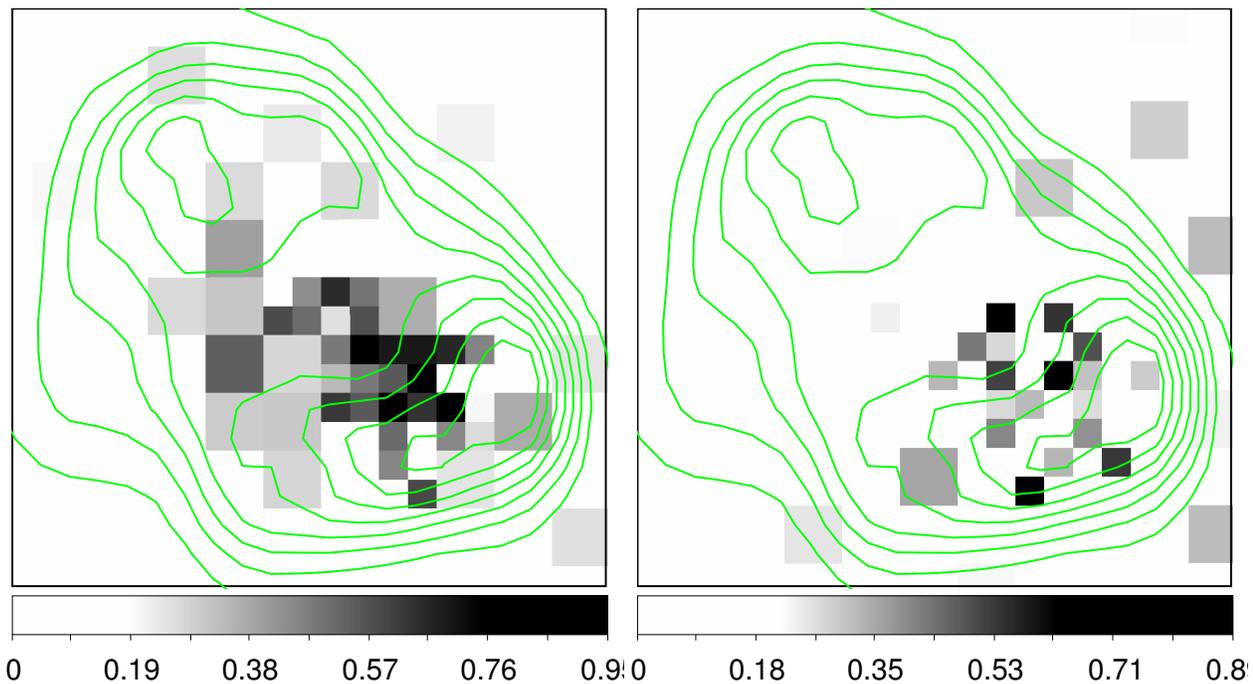}
  \caption{Equivalent width maps of the sulfur line ($\sim 2.43\keV$, {\sl left
  panel}) and the argon line ($\sim 3.16\keV$, {\sl right panel}), both overlaid
  with the same radio contours as those in Figure~\ref{fig:xmm}. The energy range we
  use to extract the sulfur line emission is 2.33--2.60$\keV$, while the
  lower- and higher-energy ranges for estimating the continuum emission at the
  line energy range are 2.27--2.30$\keV$ and 2.63--2.83$\keV$, respectively. For the argon line,
  the line energy range is 2.87--3.30$\keV$, and the lower- and higher-energy
  ranges are 2.63--2.83$\keV$ and 3.40--3.90$\keV$, respectively. The
  color bars are in linear scale in units of $\keV$.} \label{fig:ewmap}
\end{figure}

\begin{figure}[htpb]
  \plotone{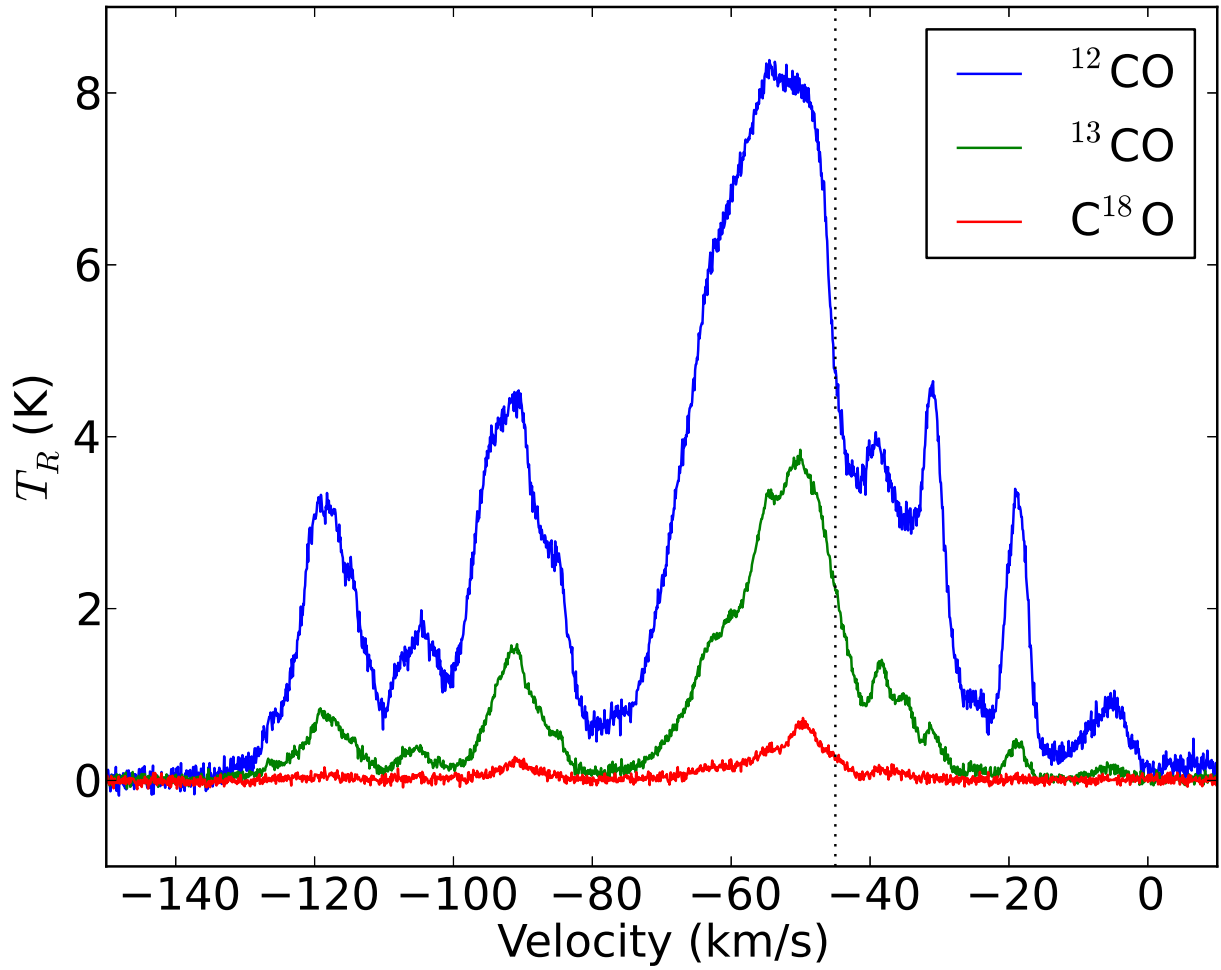}
  \caption{The spectra of \twCO~(\Jotz), \thCO~(\Jotz)\ and
  \eiCO~(\Jotz)\
  line emission in the FOV. The black dotted vertical line
  denotes the LSR velocity of the OH maser emission.}
  \label{fig:co_spec}
\end{figure}

\onecolumn
\begin{figure}[htpb]
  \plotone{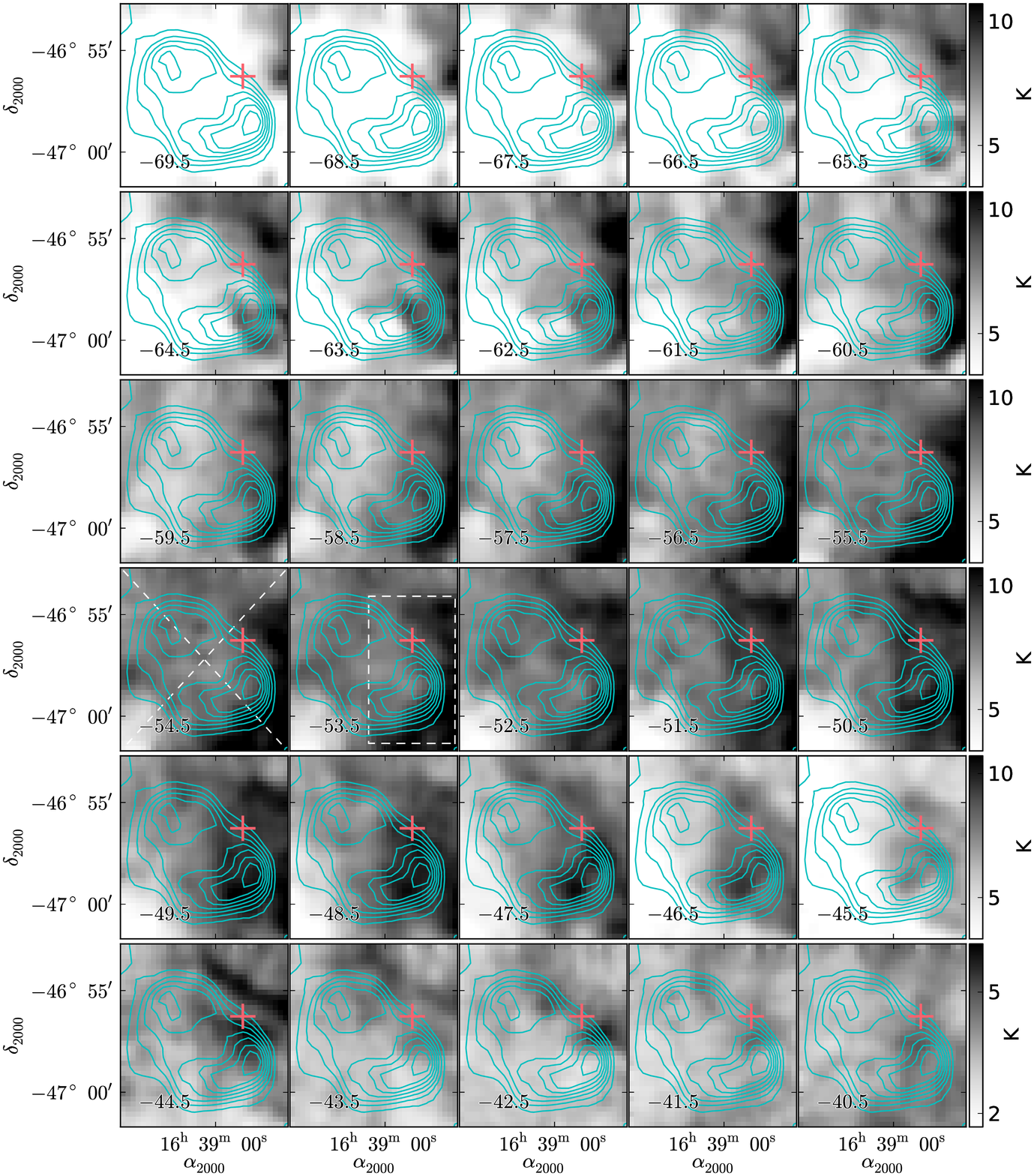}
  \caption{\small \twCO~(\Jotz) intensity maps integrated each $1\kms$
  in velocity range $-70$--$-40\kms$ in linear scale, overlaid with the MOST $843 \MHz$\
  radio continuum contours (at 7 linear scale levels between
  0.07 and 0.88$\Jyperb$). The cross indicates the location of the OH
  (1720$\MHz$) maser spot.
  The colorbar of the last row of panels is different from those of
  the other rows due to the relatively low brightness.
  The dashed lines
  in the $-54.5\km\ps$ panel
  depict the two diagonals of the FOV 
  along which the column density distribution
  $N({\rm H}_2)$ of the molecular gas in velocity range
  $-70$--$-40\kms$ are measured.
  The
  intersection of the diagonals is used as the reference point (see 
  Figure~\ref{fig:col}).
  The dashed box in the $-53.5\km\ps$ panel decipts the region we derive the parameters 
  of the interacting MC.}
  \label{fig:ch_12}
\end{figure}

\begin{figure}[htpb]
  \plotone{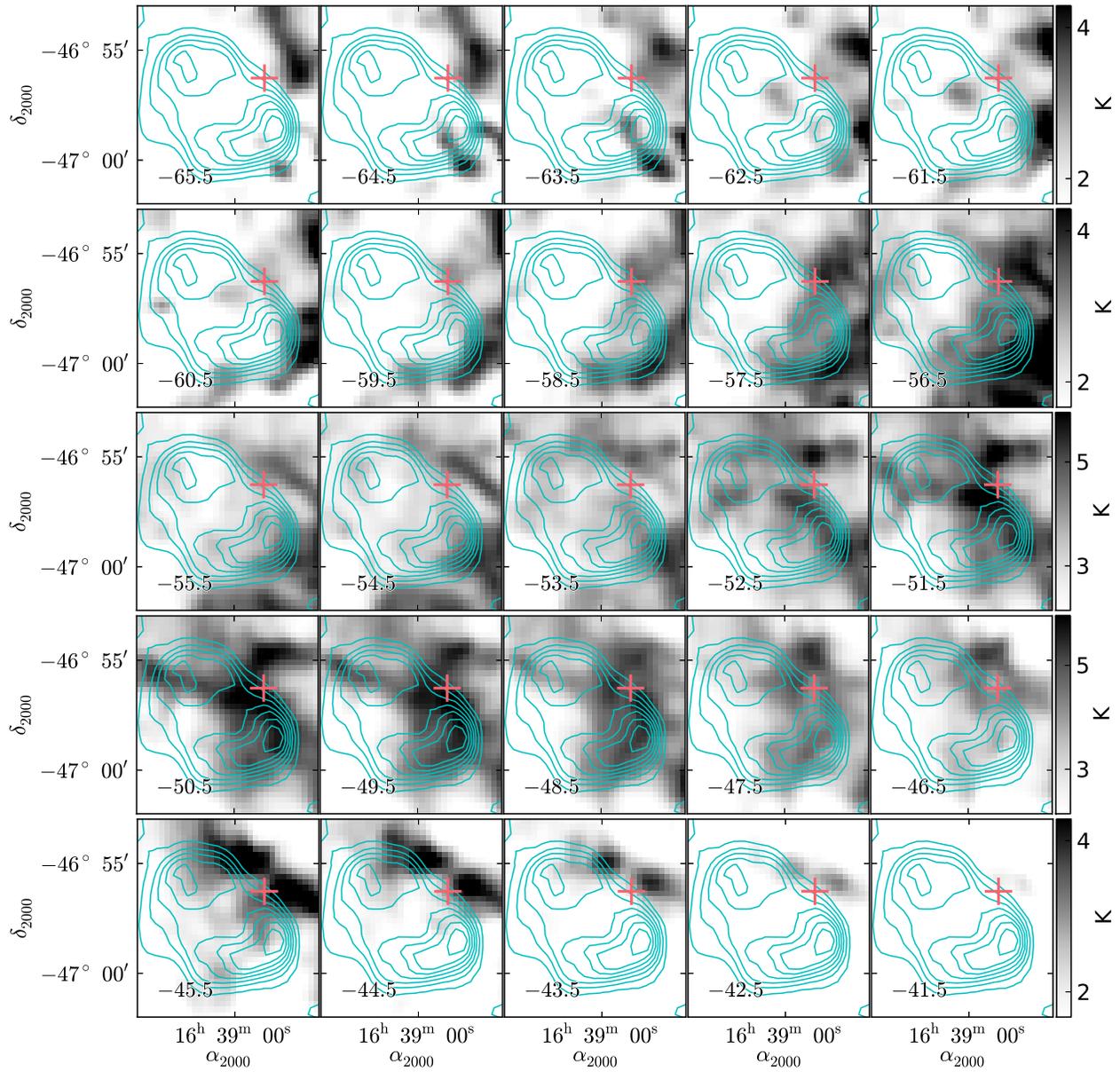}
  \caption{Intensity maps of \thCO~(\Jotz)\ observation between
  $-66\kms$\ and $-41\kms$ in linear scale. The contours and cross are the
  same as those in Figure~\ref{fig:ch_12}. 
  Note the difference in colorbars for different rows.
  }
  \label{fig:ch_13}
\end{figure}

\begin{figure}[htpb]
  \plotone{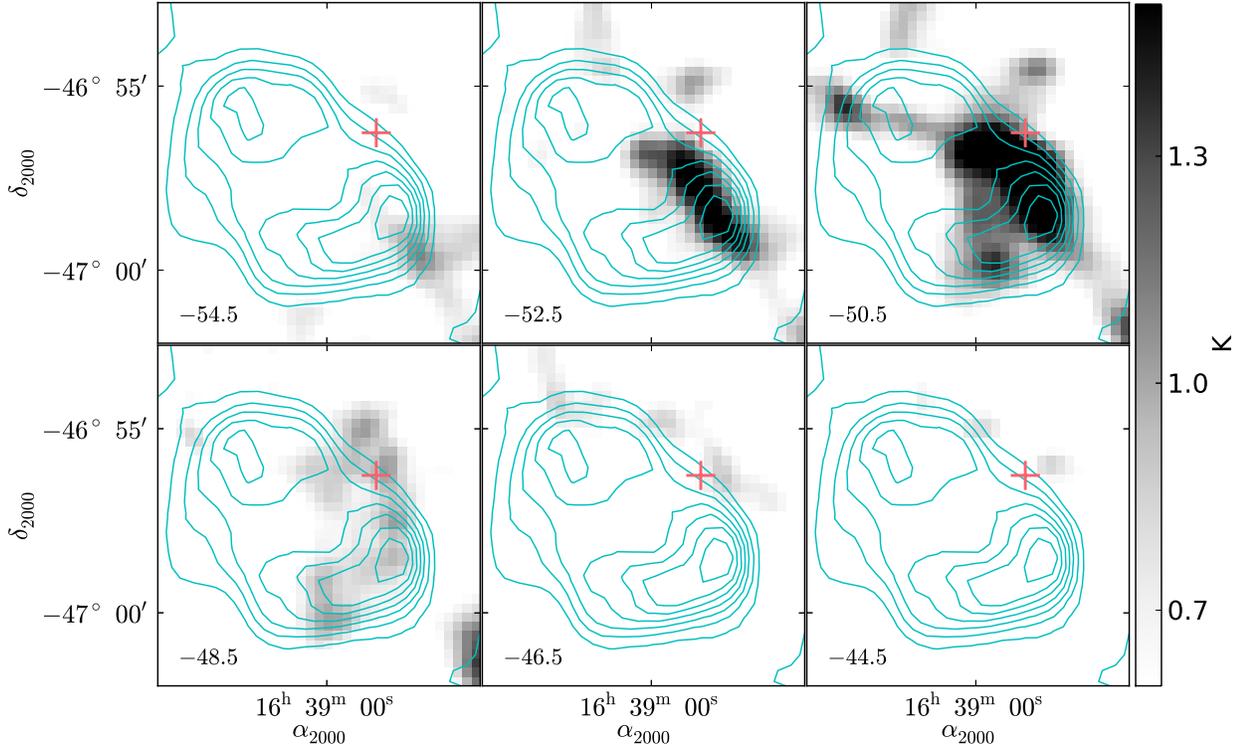}
  \caption{Intensity maps of \eiCO~(\Jotz)\ observation between
  $-55\kms$\ and $-45\kms$ in linear scale. The contours and cross are the
  same as those in Figure~\ref{fig:ch_12}.
}
  \label{fig:ch_18}
\end{figure}

\begin{figure}[htpb]
  \plotone{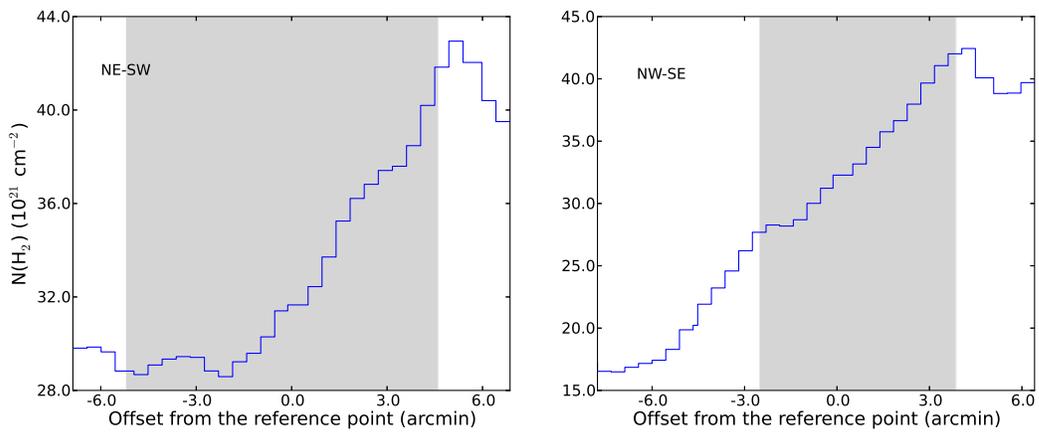}
  \caption{The column density distribution along the dashed lines illustrated in
  the $-54.5\km\ps$ panel of Figure~\ref{fig:ch_12}, the left panel corresponding
  to the northeast-southwest oriented line and the right panel 
  to the southeast-northwest oriented line. The reference point is
  the intersection of the lines. The shaded parts
  indicates the range within the SNR's radio shell.} \label{fig:col}
\end{figure}

\begin{figure}[htpb]
  \plotone{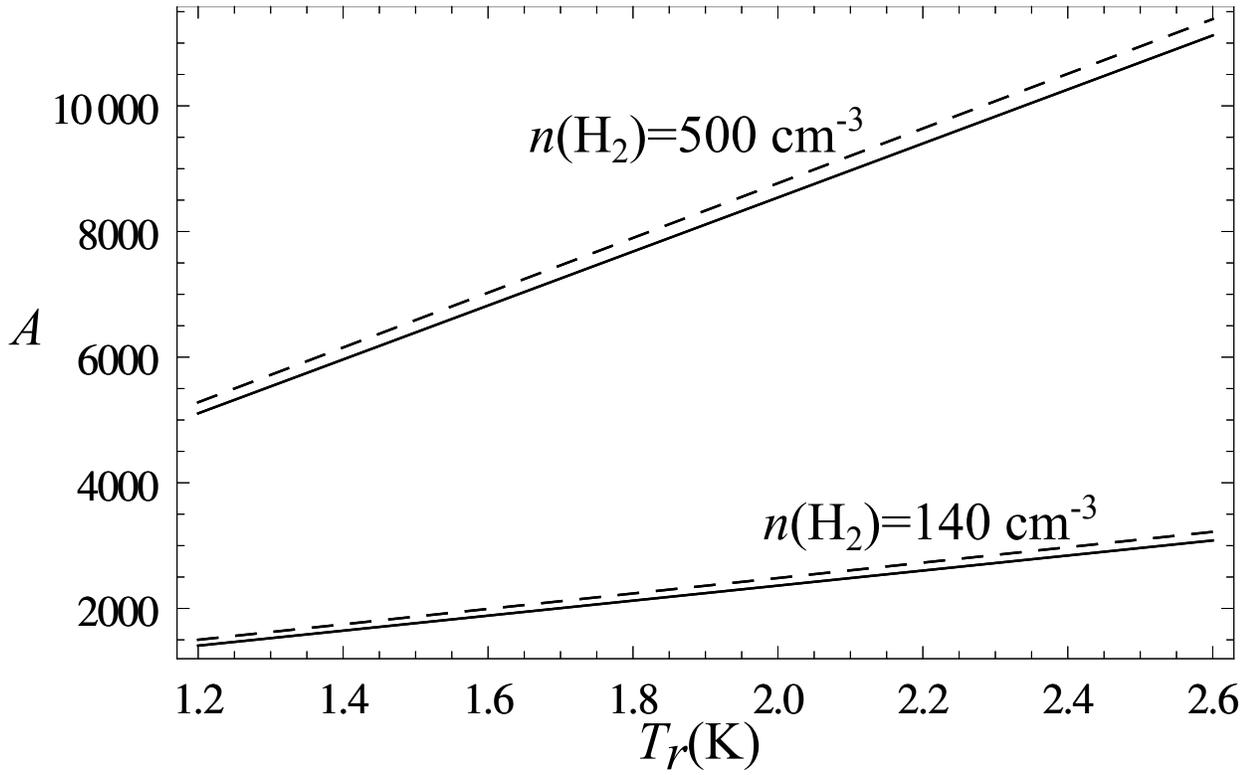}
  \caption{The relation between the wall-to-cavity contrast and the temperature
  after the heating by the reflected shock. The calculation for two exemplified molecular
  densities is given. There are two sets of solutions of Equation~(\ref{equ:ar}) for
  each density.
 } \label{fig:ref}
\end{figure}

\onecolumn
\begin{deluxetable}{llllcccc}
\tabletypesize{\tiny}
\rotate
\tablecolumns{7} 
\tablewidth{0pt} 
\tablecaption{
List of Known \MMSNR s\label{tab:mm}
}
\tablehead{  \colhead{} & \colhead{Enriched Metal} & \colhead{Temperature}&
\colhead{ionization parameter}&
  \colhead{MC} & \colhead{Pre-existing} & \colhead{Progenitor's} \\
 \colhead{SNR Name}& \colhead{Species} & \colhead{$kT_e$ (keV)\tablenotemark{a}}& 
 \colhead{$n_e t$ (s$\cm^{-3}$)\tablenotemark{b}}&
\colhead{Interaction\tablenotemark{c}}& \colhead{Cavity/Shells (Radius/pc)} &
\colhead{Mass ($\Msun$)}
}
\startdata
G0.0+0.0 (Sgr A East)     & S, Ar, Ca, Fe, Ni [1,2,3,4] &
    $kT_1\sim$1, & 
	 & Y, OH & Y ($\sim5$) [5] & 
	13--20 [1,2,3] ,$\ga12$\tablenotemark{d}\\
	& & $kT_2\sim$5 [3,2,4]\\
\hline
G6.4--0.1 (W28)           & -- & 
	$kT_1\sim$0.4,& $(n_e t)_1\sim1$--$4\E{11}$ & Y, OH & \\
	&&$kT_2\sim$0.8 [6,7,8] & [6,7,8] \\
 \emph{R}:                & -- & $\sim$0.6 [6,9] & $\sim9\E{11}$ [6,9] \\
\hline
G21.8--0.6 (Kes~69)       & -- & $\sim0.8$ [10,11,12] & & Y, OH & Y ($\sim13$) [13] & $\ga18$ [14] \\
\hline
G31.9+0.0 (3C~391)        & - & $\sim0.5$--$0.8$ [15,16,8] &
    $\ga10^{12}$ [15,8] & Y, OH & Y ($\sim5$) [18] & $\ga12$\tablenotemark{d} \\
 \emph{R}:                & Mg? [17] & $\sim0.5$--$0.6$ [17] & $\sim10^{12}$ [17] \\
\hline
G33.6+0.1 (Kes~79)        & S, Ar [19,20] & 
	$kT_1\sim0.2$, & & Y? & Y ($\sim8$) [23,20] & $\ga14$ [20] \\
	&&$kT_2\sim1.1$ [20,21,19,22] & $(n_e t)_2\sim3.9\E{10}$ \\
	&&& [20,21,19,22]\\
\hline
G34.7--0.4 (W44)		  & Ne, Si, S [24]  & 
    $kT_1\sim0.3$--0.8,  &
	$\ga3\E{12}$ [24,8,25] & Y, OH & Y ($\sim30$) [26] & 8--15? [27]\\
	& & $kT_2\sim0.9$--4 [24,8,28,25,27] \\
 \emph{R}:                &  Mg, Si, S, Ar, Ca [29] & 
	$kT_1\sim0.5$, & $6\E{11}$ [29] \\
	& & $kT_2\sim1.0$ [29]\\
\hline
G38.7--1.4                & O, Ne [30] & $\sim0.6$ [30] & $\sim7\E{12}$ [30]\\
\hline
G41.1--0.3 (3C~397)		  & O, S, Ca, Fe [31,32,33,34,35] & 
	$kT_1\sim0.21$--$0.25$, & 
	$(n_e t)_1\sim3\E{13}$,
	& Y & Y ($\sim6$) [36] & $\ga12$ [14] \\
	&& $kT_2\sim1.5$--3.5 [31,35,34] & $(n_e t)_2\sim2\E{11}$ [31,35,34]\\
\hline
G43.3--0.2 (W49B)         &  Si, S, Ar, Ca, Fe, Ni [8,37,38,39,40] &
	$kT_1\sim0.7$--1.05, & & Y? & Y? ($\sim7$) [23] & $\ga13$ [23]\\
	&& $kT_2\sim1.75$--3.3 \\
	&& [39,41,42,38,8,40]\\
  \emph{R}:               &  Si, S, Ar, Ca, Fe, Ni [43,44,45] &
	$kT_1\sim0.13$--$0.3$,  & & & & $\sim25$ [37,39,41,40]\\
	&& $kT_2\sim1.12$--1.91 [43,44,45]\\
\hline
G49.2--0.7 (W51C)         & Ne [46], Mg [46,47], S [48] & 
	0.56--0.74 [46,47,49,48] & 0.8--11$\E{11}$ & Y, OH & & $\ga20$ [46] \\
\hline
G53.6--2.2 (3C~400.2)     & Fe? [50] & $\sim0.8$ [50] & $\sim10^{11}$ [50] \\
\hline
G65.3+5.7                 & -- &$\sim0.25$ [51]&\\
\hline
G82.2+5.3 (W63)			  & Mg, Si, Fe [52]  & 
	$kT_1\sim0.2$, &  \\
	&& $kT_2\sim0.6$ [52,53]\\
\hline
G85.4+0.7                 & O? [54] & 1.0 [54] & 8$\E{10}$ [54] \\
\hline
G85.9--0.6                & O, Fe [54] & 1.6 [54] & 5$\E{10}$ [54] \\
\hline
G89.0+4.7 (HB21)		  & Si, S [55,56] & 0.62--0.68 [56,55] & $\ga4\E{11}$ [55,56] & Y & \\
\hline
G93.3+6.9 (DA~530)        & Si [57] & 0.3--0.6 [57,58] & $>4\E{11}$ [57]& & & $\sim10$ [57] \\
\hline
G93.7-0.2 (CTB~104A)? [53] & -- & & &\\
\hline
G116.9+0.2 (CTB 1)		  & Mg [55], O, Ne [56] & 
	$kT_1\sim0.2$--0.3,  & $\ga1\E{11}$ [56,55]& & & $\ga13$ [56] \\
	&& $kT_2\sim3$ [56],\\
	&& or $kT_2\sim0.8$ [55]\\
\hline
G132.7+1.3 (HB~3)		  & Mg [55], or Mg, Ne, O [55] &
	$\sim0.3$ [55, 53] && Y? & Y ($\sim70$) [59] & \\
\hline
G156.2+5.7                & Si, S [60,61,62] &
	$kT_1\sim0.45$, & $\sim1.5\E{11}$ [60,61,62] &  & & $\la15$ [61] \\
	&& $kT_2\sim0.6$ [62,60,63,61]\\
\hline
G160.9+2.6 (HB~9)         & -- & $\sim0.8$ [64] && ? \\
\hline
G166.0+4.3 (VRO 42.05.01) & S [65] & $\sim0.7$ [66,67]&& ? & \\
\hline
G189.1+3.0 (IC443)        & Mg , S [68,69,65], Si [68,69], Ne [65] &
	$kT_1\sim0.3$--0.7, &
	$(n_e t)_1\ga10^{12}$ [68,69]& Y, OH & Y ($\sim11$) [75] & $\ga15$ [75] \\
	&& $kT_2\sim1.0$--1.8 \\
	&& [68,69,65,70,71,53]\\
   \emph{R}:              & Si, S, Ar [72], Ca, Fe, Ni [73] &
	$\sim0.65$ [72,73,74] & $\sim9.8\E{11}$ [73] \\
\hline
G272.2-3.2                & O [76], Ne [77], Si, S, Fe [76,77,78], Ca, Ni [78] &
	$\sim0.7$--1.5 [77,76,78,79,80] & 2--10$\E{10}$ [77,76,78,80] & & & (SN Ia) [78,77,76]\\
\hline
G290.1--0.8 (MSH 11--61A) & Si, S [81,82], Mg [81] &
	$\sim0.6$--0.9 [82,81] & $>0.1\E{13}$ [82] & ? & & 25--30 [82] \\
\hline
G304.6+0.1 (Kes~17)       & Mg? [83,84,85] & $\sim0.79$--$1.10$ [84,86,85,83] &
	$\ga3.7\E{11}$ [84,86,85,83] & Y & \\
\hline
G311.5-0.3                & -- &$\sim0.98$ [84]& & ? \\
\hline
G327.4+0.4 (Kes~27)       & S, Ca [87], Si [8] & 
	$\sim0.5$--1.2 [87,53,88,8,89] & $\ga1\E{11}$ [87,8]&  & Y [90] \\
\hline
G337.8--0.1 (Kes~41)      & S, Ar [91] & $\sim$1.9 [92,91] &
	$\sim2\E{11}$ [92,91] & Y, OH & Y ($\sim13$) [91] & $\ga18$\tablenotemark{d} \\
\hline
G344.7--0.1               & Al, Si, S, Ar, Ca, Fe [93,94,95,96] &
	$\sim$0.8--1.8 [96,95,94,93] & 1--4$\E{11}$ [96,95,94,93] & ? & & (SN Ia) [93]\\
\hline
G346.6-0.2                & Ca [97] & $\sim1.2$ [97,98,84] & $\sim2.9\E{11}$ [97,84] & Y, OH & \\
 \emph{R}:                & --  & $\sim$0.3 [99]& $\sim4.8\E{11}$ [99] \\
\hline
G348.5+0.1 (CTB~37A)      & -- & 0.55--0.83 [84,98,100,101,102] &
	$>3\E{10}$ [84,102] & Y, OH & ? [103] \\
  \emph{R}:               & Si? [102] & $\sim$0.5 [102] & $\sim1.3\E{12}$ [102] &\\
\hline
G352.7--0.1               & Si, S [104,105,19], Ar? [19,104] & 
	$0.8$--2.1 [104,105,4] & 1--2$\E{10}$ [104,105,19] &  \\
\hline
G355.6--0.0               & Si, S, Ar, Ca [106] & $\sim$0.6 [106,98]&  \\
\hline
G357.7-0.1 (Tornado)      & -- & $\sim$0.6 [11,107]&& Y,OH\\
\hline
G359.1--0.5               & Si, S [108,109] & 
	$kT_1\sim1.1$,  & & Y, OH & Y ($\sim28$) [110] & \\
	&& $kT_2\sim2.0$ [109,111,108,112]\\
   \emph{R}:              & Si, Mg, S [113] & $\sim0.3$ [113] \\
\enddata 
\onecolumn
\tablecomments{``Y" in the table means ``yes"; 
``?" means that the property is uncertain or
controversial;
``\emph{R}" denote that the parameters are derived using the recombining
plasma model for the same SNR.
}
\tablenotetext{a}{The different tmperatures of different regions are shown as a range, and 
``$kT_1$, $kT_2$" are temperatures of cold and hot components respectively.}
\tablenotetext{b}{The ionization parameter of underionized state and
overionized state respectively.}
\tablenotetext{c}{This column is adopted from Jiang et al.'s (2010)
SNR-MC association table.
``OH" means that the 1720~MHz OH maser is detected.
}
%
%
\tablenotetext{d}{Estimated using the
method described in \citet{Chen2013}. This method is not applied to 
SNRs W44, HB~3, and G359.1$-$0.5 which are also in molecular cavities,
considering the cavity sizes are larger than the scope
of application.}
%
%
\tablerefs{[1] \citet{Maeda2002}; [2] \citet{Sakano2004}; [3] \citet{Park2005};
[4] \citet{Koyama2007}; [5] \citet{Lee2008}; [6] \citet{Zhou2014};
[7] \citet{Rho2002}; [8] \citet{Kawasaki2005}; [9] \citet{Sawada2012};
[10] \citet{Bocchino2012}; [11] \citet{Yusef2003}; [12] \citet{Seo2013};
[13] \citet{Zhou2009}; [14] \citet{Chen2013};
[15] \citet{Chen2004}; [16] \citet{Chen2001}; [17] \citet{Ergin2014};
[18] \citet{Reach2002};
[19] \citet{Giacani2009}; [20] Zhou et al. in preparation;
[21] \citet{Sun2004}; [22] \citet{Auchettl2014}; [23] \citet{Chen2014};
[24] \citet{Shelton2004};
[25] \citet{Harrus1997}; [26] \citet{Seta2004}; [27] \citet{Rho1994};
[28] \citet{Harrus2006}; [29] \citet{Uchida2012}; [30] \citet{Huang2014};
[31] \citet{Safi2005}; [32] Safi-Harb et al. in preparation; [33] \citet*{Jiang2010SC};
[34] \citet{Chen1999}; [35] \citet{Safi2000}; [36] \citet{Jiang2010}; 
[37] \citet{Keohane2007}; [38] \citet{Hwang2000};
[39] \citet{Miceli2006}; [40] \citet{Lopez2013a}; [41] \citet{Lopez2009};
[42] \citet{Fujimoto1995}; [43] \citet{Lopez2013b}; [44] \citet{Ozawa2009};
[45] \citet{Miceli2010}; [46] \citet{Sasaki2014}; [47] \citet{Hanabata2013};
[48] \citet{Koo2005}; [49] \citet{Koo2002}; [50] \citet{Yoshita2001};
[51] \citet{Shelton2004b}; [52] \citet{Mavromatakis2004}; [53] \citet{Rho1998};
[54] \citet{Jackson2008}; [55] \citet{Lazendic2006}; [56] \citet{Pannuti2010};
[57] \citet{Jiang2007}; [58] \citet{Kaplan2004}; [59] \citet{Routledge1991};
[60] \citet{Yamauchi1999}; [61] \citet{Katsuda2009}; [62] \citet{Uchida2012b};
[63] \citet{Pannuti2004}; [64] \citet{Leahy1995}; [65] \citet{Bocchino2009};
[66] \citet{Burrows1994}; [67] \citet{Guo1997}; [68] \citet{Troja2006};
[69] \citet{Troja2008}; [70] \citet{Petre1988}; [71] \citet{Keohane1997};
[72] \citet{Yamaguchi2009}; [73] \citet{Ohnishi2014}; [74] \citet{Kawasaki2002};
[75] \citet{Su2014}; [76] \citet{Sanchez2013}; [77] \citet{McEntaffer2013}; 
[78] \citet{Sezer2012};
[79] \citet{Greiner1994}; [80] \citet{Harrus2001}; [81] \citet{Slane2002}; 
[82] \citet{Garcia2012}; [83] \citet{Gelfand2013}; [84] \citet{Pannuti2014b};
[85] \citet{Gok2012}; [86] \citet{Combi2010b}; [87] \citet{Chen2008};
[88] \citet{Enoguchi2002}; [89] \citet{Seward1996}; [90] \citet{McClure2001};
[91] this paper;  [92] \citet{Combi2008}; [93] \citet{Yamaguchi2012};
[94] \citet{Giacani2011}; [95] \citet{Yamauchi2005}; [96] \citet{Combi2010};
[97] \citet{Sezer2011}; [98] \citet{Yamauchi2008}; [99] \citet{Yamauchi2013};
[100] \citet{Aharonian2008b}; [101] \citet{Sezer2011b}; [102] \citet{Yamauchi2014};
[103] \citet{Maxted2013}; [104] \citet{Pannuti2014}; [105] \citet{Kinugasa1998};
[106] \citet{Minami2013}; [107] \citet{Gaensler2003}; [108] \citet{Bamba2000};
[109] \citet{Bamba2009}; [110] \citet{Uchida1992};
[111] \citet{Egger1998}; [112] \citet{Sakano2002};
[113] \citet{Ohnishi2011}; 
}
\end{deluxetable}
\end{document}